\documentclass[conference]{IEEEtran}
\IEEEoverridecommandlockouts

\usepackage{amsmath,amssymb,amsfonts}
\usepackage{algorithmic}
\usepackage{graphicx}
\usepackage{textcomp}
\usepackage{xcolor}
\usepackage{algorithm}
\usepackage{braket}
\usepackage{orcidlink}

\usepackage[style=ieee, backend=biber]{biblatex}
\begin{document}
\title{DQAOA-GPT: AI-Accelerated Distributed Quantum Optimization for Combinatorial Problems

\thanks{This research used resources of the Oak Ridge Leadership Computing Facility at the Oak Ridge National Laboratory, which is supported by the Office of Science of the U.S. Department of Energy under Contract No. DE-AC05-00OR22725. 
{\it Notice}: This manuscript has in part been authored by UT-Battelle, LLC under Contract No. DE-AC05-00OR22725 with the U.S. Department of Energy. The United States Government retains and the publisher, by accepting the article for publication, acknowledges that the U.S. Government retains a non-exclusive, paid up, irrevocable, world-wide license to publish or reproduce the published form of the manuscript, or allow others to do so, for U.S. Government purposes. The Department of Energy will provide public access to these results of federally sponsored research in accordance with the DOE Public Access Plan (http://energy.gov/downloads/doe-publicaccess-plan).}
}

\author{
\IEEEauthorblockN{
Seongmin Kim\,\orcidlink{0000-0001-5906-3004}\IEEEauthorrefmark{1},
Abhinav Rijal\,\orcidlink{0000-0002-5551-6595}\IEEEauthorrefmark{2},
Yuri Alexeev\,\orcidlink{0000-0001-5066-2254}\IEEEauthorrefmark{3},
Nora Bauer\,\orcidlink{0000-0002-8723-4656}\IEEEauthorrefmark{4}, \\
Martin Roetteler\,\orcidlink{0000-0003-0234-2496}\IEEEauthorrefmark{5},
Mina Yoon\,\orcidlink{0000-0002-1317-3301}\IEEEauthorrefmark{6},
George Siopsis\,\orcidlink{0000-0002-1466-2772}\IEEEauthorrefmark{4},
In-Saeng Suh\,\orcidlink{0000-0002-6923-6455}\IEEEauthorrefmark{1}
}

\IEEEauthorblockA{\IEEEauthorrefmark{1}
National Center for Computational Sciences, Oak Ridge National Laboratory,
Oak Ridge, TN, USA
}
\IEEEauthorblockA{\IEEEauthorrefmark{2}
Department of Physics \& Astronomy, University of Tennessee,
Knoxville, TN, USA
}
\IEEEauthorblockA{\IEEEauthorrefmark{3}
NVIDIA Corporation, Santa Clara, CA, USA
}
\IEEEauthorblockA{\IEEEauthorrefmark{4}
IonQ Inc., Chattanooga, TN, USA
}
\IEEEauthorblockA{\IEEEauthorrefmark{5}
IonQ Inc., Seattle, WA, USA
}
\IEEEauthorblockA{\IEEEauthorrefmark{6}
Materials Science and Technology Division, Oak Ridge National Laboratory,
Oak Ridge, TN, USA
}
}

\maketitle

\begin{abstract}
While combinatorial optimization problems are central to many scientific and engineering applications, their solution remains challenging due to exponentially large search spaces. Variational quantum algorithms offer a promising route for tackling such problems, yet their practical performance is limited by repeated quantum circuit evaluations and classical parameter updates. In this work, we introduce DQAOA-GPT, a hybrid framework that integrates the distributed quantum approximate optimization algorithm (DQAOA), which decomposes a large optimization problem into smaller sub-problems, with GPT-based quantum circuit generation for solving those sub-problems. Rather than relying on iterative variational optimization, the proposed approach uses a trained generative model to directly generate high-quality quantum circuits for the decomposed sub-problems. As a benchmark, we evaluate DQAOA-GPT against conventional DQAOA on dense HUBO optimization problems with up to 100 decision variables. The results demonstrate that DQAOA-GPT significantly reduces computational cost while maintaining competitive solution quality, with larger acceleration observed for larger sub-problem sizes. Although this work focuses on benchmark-scale validation, the framework provides a promising foundation for larger-scale combinatorial optimization in hybrid HPC-QC environments through increased GPU resources and parallel computing capability.
\end{abstract}


\section{Introduction}

Combinatorial optimization problems arise in a wide range of scientific and engineering applications, including finance, logistics, networks, and materials design \cite{brandhofer2022benchmarking, ushijima2021multilevel, lykov2023sampling, kim2026harnessing}. These problems are characterized by large discrete search spaces, complex variable interactions, and highly non-convex objective landscapes, making them difficult to solve efficiently with conventional optimization methods \cite{kitai2020designing, kim2022high, kim2025quantum}. 

Quantum computing (QC) has attracted considerable attention as a potential computational paradigm for combinatorial optimization \cite{cerezo2021variational, sankar2024benchmarking, bauer2024combinatorial}. Among the most widely studied approaches, quantum approximate optimization algorithm (QAOA) uses parameterized quantum circuits to prepare approximate ground states of problem Hamiltonians through alternating cost and mixer evolutions \cite{farhi2014quantum, shaydulin2024evidence}. Although QAOA has emerged as a leading candidate for near-term quantum optimization, standard QAOA faces major limitations when applied to large-scale combinatorial optimization problems. As the problem size increases, direct encoding requires more qubits, deeper circuits, and repeated quantum circuit evaluations during classical parameter optimization, where the optimization landscape is more complex and the variational loop becomes increasingly expensive.

To address these limitations, distributed QAOA (DQAOA) has been developed as a hybrid high-performance computing (HPC)-QC framework for solving large optimization problems through problem decomposition and coordinated execution \cite{kim2024distributed, xu2025gpu}. This HPC-QC integration improves computational scalability and enables large problems that are difficult to handle with standard QAOA. Despite this advantage, DQAOA still relies on iterative variational optimization for each decomposed sub-problem, requiring repeated quantum circuit executions and classical parameter updates. As a result, the variational loop remains a major computational bottleneck, particularly when the decomposed sub-problem size becomes large.

Recent advances in generative AI, particularly transformer architectures, have opened new directions for quantum algorithm design \cite{nakaji2024generative, alexeev2025artificial, tyagin2025qaoa}. QAOA-GPT aims to learn a direct mapping from problem instances to high-quality quantum circuits, bypassing iterative variational optimization \cite{tyagin2025qaoa, sunny2025extending}. By generating optimized circuit structures at inference time, these methods can significantly reduce the quantum-classical feedback loop and improve overall computational efficiency. Nevertheless, existing AI-assisted approaches are primarily demonstrated on small- to moderate-scale problem sizes and do not explicitly address the scalability challenges associated with large-scale combinatorial optimization problems.

In this work, we introduce DQAOA-GPT, a hybrid computational framework designed to address large-scale combinatorial optimization by tightly integrating distributed quantum optimization with AI-driven circuit generation \cite{kim2024distributed, tyagin2025qaoa}. The framework consists of two key components:
(i) DQAOA-based decomposition and aggregation strategy that enables scalable execution across HPC–QC environments, and
(ii) GPT-based circuit generation model that replaces iterative parameter optimization with direct circuit synthesis. 

By coupling distributed execution with learned circuit generation, DQAOA-GPT eliminates one of the primary bottlenecks in variational quantum algorithms while preserving the scalability advantages of DQAOA. This integration enables efficient exploration of large and complex optimization landscapes with reduced computational overhead.
 
The contributions of this work are:
\begin{itemize}
\item We propose DQAOA-GPT, a unified framework integrating distributed quantum optimization with generative circuit synthesis.
\item We employ a GPT-based approach that removes the need for variational parameter optimization, reducing the computational overhead of repeated quantum circuit evaluations.
\end{itemize}

Overall, this work highlights the potential of combining HPC, AI, and QC to build scalable hybrid optimization frameworks and move toward practical quantum utility.

\section{Background}

\subsection{Combinatorial Optimization Problem}

Combinatorial optimization problems aim to identify an optimal configuration from a discrete set of candidate solutions. In many scientific and engineering applications, such problems can be formulated using binary decision variables and represented as polynomial objective functions \cite{kitai2020designing, kim2022high}. In this work, we consider a higher-order unconstrained binary optimization (HUBO) form, which provides a flexible representation for combinatorial optimization problems involving complex interactions among variables.

A third-order HUBO problem is defined as a real-valued polynomial objective over binary variables $\mathbf{x} \in {0,1}^N$ \cite{hwang2025higher}:
 
\begin{equation}
    \begin{aligned}
        \min_{\mathbf{x} \in \{0,1\}^N} \; H(\mathbf{x}) 
        &= \sum_{i} c_i\, x_i
        + \sum_{i < j} c_{ij}\, x_i x_j
        + \sum_{i < j < k} c_{ijk}\, x_i x_j x_k 
    \end{aligned}
\label{eq:hubo}
\end{equation}

where $c_i$, $c_{ij}$, and $c_{ijk}$ denote linear, quadratic, and cubic interaction coefficients. Solving HUBO problems is NP-hard in general \cite{mandal2020compressed}. This optimization problem can be compactly represented using a tensor $T \in \mathbb{R}^{N \times N \times N}$ in upper-triangular form, where entries $T[i,i,i]$ encode linear terms, $T[i,i,j]$ ($i < j$) encode quadratic interactions, and $T[i,j,k]$ ($i < j < k$) encode cubic interactions.
 
For quantum optimization, binary variables are mapped to spin variables via $x_i = (1 - z_i)/2$, converting the objective into an Ising Hamiltonian with Pauli-$Z$ operators. Higher-order terms map to multi-qubit operators (e.g., $Z_i Z_j Z_k$), allowing direct encoding without quadratization \cite{glos2022space}.

In this work, we consider HUBO instances motivated by materials optimization applications, with a problem size of $N = 100$. To evaluate solution quality, we use the best-known results reported in Ref. \cite{kim2026distributed} as reference solutions and calculate the relative accuracy accordingly.

\subsection{ADAPT-QAOA}
 
ADAPT-QAOA constructs a problem-adaptive ansatz from a predefined operator pool $\mathcal{P} = {O_1, \ldots, O_M}$ \cite{zhu2022adaptive}. At iteration $k$, an operator $O^{(k)} \in \mathcal{P}$ is selected based on the energy gradient magnitude:
 
\begin{equation}
    g_j = \frac{\partial E}{\partial \varepsilon_j}
        = -i\,\bigl\langle\psi^{(k-1)}\bigr|
          e^{i\gamma_0 H_c}
          \bigl[H_c,\, O_j\bigr]
          e^{-i\gamma_0 H_c}
          \bigl|\psi^{(k-1)}\bigr\rangle,          
    \label{eq:adapt_grad}
\end{equation}
 
where $\gamma_0$ is the initial gamma, $H_c$ is the cost Hamiltonian, and $\ket{\psi^{(k-1)}}$ is the previous variational state. After each operator insertion, all variational parameters are globally re-optimized. The procedure continues until a stopping criterion is satisfied \cite{zhu2022adaptive}.
 
While ADAPT-QAOA can produce high-quality circuits, it introduces substantial overhead: each iteration requires gradient evaluation across the operator pool and global parameter optimization, limiting applicability to large-scale problems.

\subsection{FEATHER Graph Embeddings}
 
FEATHER generates node representations from characteristic functions of random walk distributions on a graph \cite{rozemberczki2020characteristic}. A global descriptor is obtained by mean pooling node embeddings, producing a fixed-dimensional, permutation-invariant representation. In this work, FEATHER embeddings encode the structural information of HUBO interaction graphs for conditioning generative models.

\subsection{Generative Pre-trained Transformer}
 
GPT is a decoder-only transformer that models sequential data via autoregressive next-token prediction \cite{vaswani2017attention}. Given a token sequence $\mathbf{u} = \{u_1, \ldots,u_n\}$, the model maximizes
 
\begin{equation}
    \mathcal{L}(\mathbf{u})
    = \sum_{i=1}^{n} \log P\!\left(u_i \mid u_1, \ldots, u_{i-1};\,
      \theta\right),
    \label{eq:gpt_loss}
\end{equation}
 
where $\theta$ denotes model parameters. Causal masking ensures predictions depend only on preceding tokens, making GPT well-suited for structured sequence generation \cite{vaswani2017attention}. Quantum circuits naturally fit this paradigm, as they can be represented as sequences of gates and parameters \cite{nakaji2024generative, tyagin2025qaoa, sunny2025extending}.

\subsection{QAOA-GPT}
 
QAOA-GPT replaces the iterative optimization loop of variational quantum algorithms with generative circuit synthesis, producing a high-quality circuit through a single forward pass.
 
\paragraph{Training pipeline}
(1) sub-problem instances are solved using ADAPT-QAOA to obtain reference circuits, filtered by a target approximation ratio \cite{zhu2022adaptive}.
(2) Each sub-problem and circuit are converted into a token sequence using (index, coefficient) pairs for the problem and operator/parameter tokens for the circuit.
(3) A decoder-only transformer is trained on these sequences using next-token prediction \cite{vaswani2017attention}.
 
\paragraph{Inference}
A new sub-problem is tokenized and passed to the trained model, which autoregressively generates a complete circuit without additional optimization.
 
\begin{figure}[!ht]
    \centering
    \includegraphics[width=1.0\linewidth]{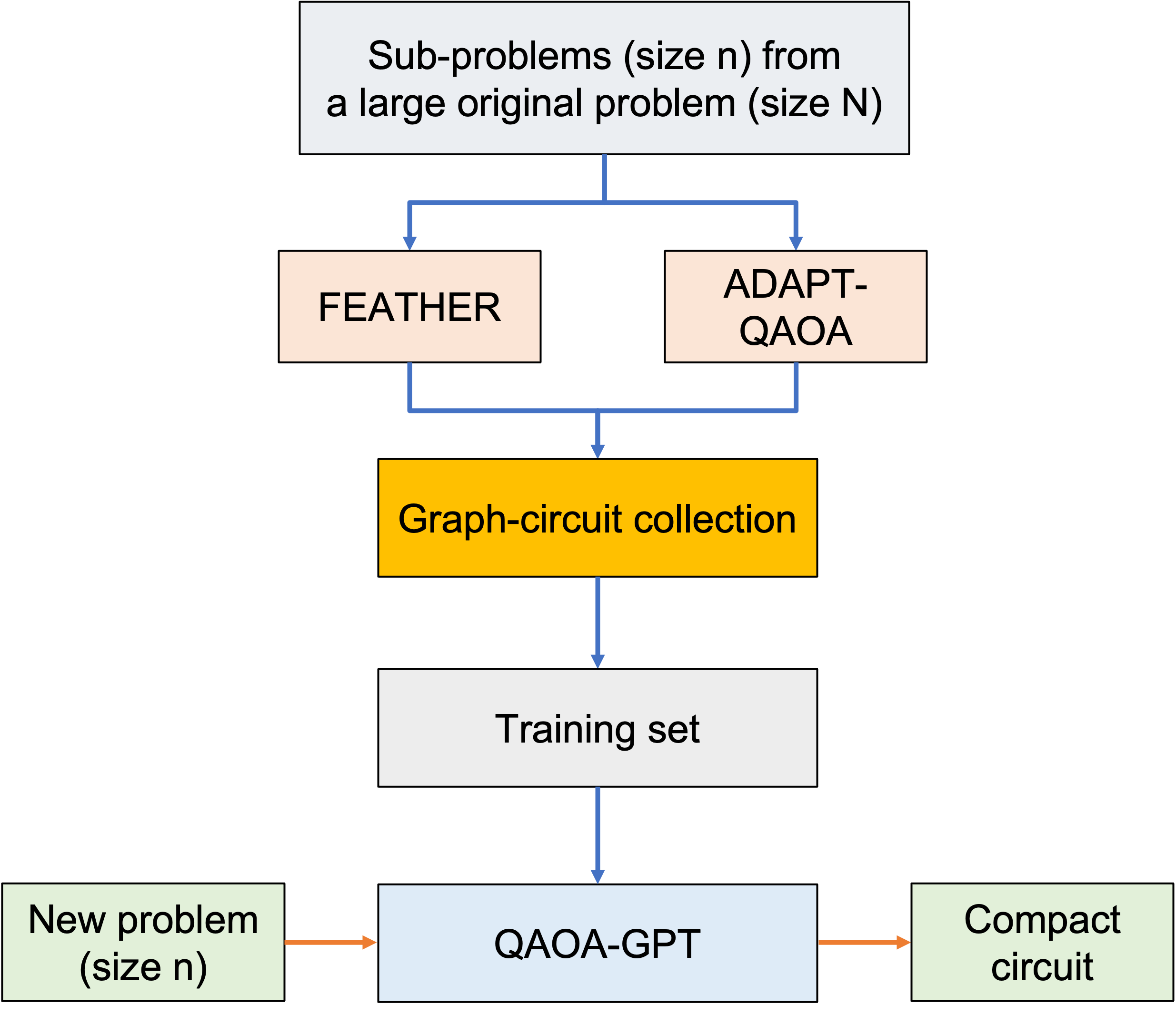}
    \caption{\label{fig:QAOA_GPT} Schematic illustration of the QAOA-GPT workflow.}
\end{figure}

QAOA-GPT produces circuits with quality comparable to ADAPT-QAOA while reducing generation time by orders of magnitude. Figure~\ref{fig:QAOA_GPT} shows the workflow. Further details are in Ref.~\cite{tyagin2025qaoa}.
 
\section{DQAOA-GPT}

We propose DQAOA-GPT, a hybrid framework for solving a large $N$-variable combinatorial optimization problem by combining iterative problem decomposition with GPT-based circuit generation for sub-problems. The overall workflow consists of two stages \cite{kim2024distributed, tyagin2025qaoa}:
(i) a training stage, in which a GPT model is trained to generate high-quality circuits for small sub-problems, and
(ii) an inference stage, in which the trained model is embedded into a DQAOA loop to iteratively improve the solution of the large original instance.

\subsection{Sub-Problem Decomposition}

Let a original optimization problem be represented by a tensor $T \in \mathbb{R}^{N \times N \times N}$ in upper-triangular form \cite{kim2026distributed}. At each DQAOA iteration, a sub-problem of size $n \ll N$ is constructed by selecting a subset of variable indices $\mathcal{S} = {s_1, \ldots, s_n} \subseteq {1,\ldots,N}$ and extracting the induced sub-tensor

\begin{equation}
    T_{\mathcal{S}}[i,j,k] = T[s_i, s_j, s_k],
    \quad i,j,k \in {1,\ldots,n}.
    \label{eq:sub_hubo}
\end{equation}

This induced sub-problem preserves all linear, quadratic, and cubic interactions among the selected variables. The resulting $n$-variable problem is then solved independently to generate a candidate local update for the global solution. In this work, the sub-problem size is chosen from $n \in {4,6,8,10,12}$, and the GPT model is trained separately for different $n$ values. 

A point worth clarifying is that the sub-problem alone does not fully determine the best local update for the global problem, because interactions with variables outside $\mathcal{S}$ remain fixed by the current global assignment. Therefore, during inference, the extracted sub-problem is interpreted as a local optimization block embedded within the current global configuration, and its candidate solution is accepted only if it lowers the global energy.

\subsection{HUBO to Graph Projection}

To incorporate structural information into the generative model, each sub-problem is projected onto a weighted graph $G=(V,E,w)$ with $|V|=n$. Each variable corresponds to a node, and interaction strengths are converted into edge weights. Quadratic terms contribute directly to pairwise edges, while cubic terms are distributed over the three edges of the corresponding triangle:
\begin{align}
    w(i,j) &\mathrel{+}= |T[i,i,j]|, \quad i < j,
    \label{eq:proj_quad} \\
    w(i,j),\,w(i,k),\,w(j,k)
    &\mathrel{+}= \tfrac{1}{3}|T[i,j,k]|, \quad i < j < k.
    \label{eq:proj_cubic}
\end{align}

This projection maps higher-order interactions onto a graph representation that approximately captures the structural coupling pattern of the sub-problem. FEATHER embeddings are then computed from this projected graph to provide a fixed-length structural descriptor for each instance \cite{rozemberczki2020characteristic}.

We emphasize that this projection is not an exact reduction of the HUBO objective to a pairwise graph. Because higher-order couplings are compressed into pairwise edge statistics, the projected graph serves only as a coarse structural summary of the sub-problem. This approximation does not remove the original higher-order information, because the exact interaction terms are provided separately in the model input. In particular, quadratic terms are represented by their indices and coefficients, $(i,j,c_{ij})$, while cubic terms are represented by $(i,j,k,c_{ijk})$. Therefore, the interaction order and coefficient information are preserved during model input construction. The FEATHER embedding derived from the projected graph is used only as an auxiliary structural descriptor for conditioning the GPT model, while the quantum circuit construction is performed using the exact interaction coefficients rather than the projected graph.

\subsection{Training the GPT model}

A large set of sub-problems is generated by random sampling from the target problem. Each sub-problem is mapped to an Ising cost Hamiltonian $H_c$ via the substitution $x_i = (1-z_i)/2$, which expands cubic terms into $Z$, $ZZ$, and $ZZZ$ Pauli contributions with coefficients scaled by $1/8$. These Hamiltonians are then solved using ADAPT-QAOA to obtain high-quality reference circuits \cite{zhu2022adaptive, glos2022space}.

Only reference circuits meeting a predefined quality threshold are retained for training, ensuring that the model learns from near-optimal examples. This procedure produces paired data consisting of a sub-problem instance and a corresponding adaptive quantum circuit \cite{tyagin2025qaoa, sunny2025extending}.

Each training instance is serialized into a token sequence. The vocabulary extends the original QAOA-GPT representation to support third-order combinatorial optimization structure:

\begin{itemize}
    \item \textbf{Special tokens:} \texttt{<bos>},
          \texttt{<end\_of\_problem>}, \texttt{<new\_layer>},
          \texttt{<eos>}, \texttt{<pad>}.
    \item \textbf{Problem tokens:} one token per distinct interaction
          index tuple --- 1-tuples $(i)$ for linear terms, 2-tuples
          $(i,j)$ for quadratic terms, and 3-tuples $(i,j,k)$ for
          cubic terms. For $n$-node sub-problems this yields
          $n + \binom{n}{2} + \binom{n}{3}$ distinct index tokens.
    \item \textbf{Coefficient tokens:} a shared grid of 201 values
          covering $[-10, 10]$ in steps of $0.1$.
    \item \textbf{Circuit tokens:} one token per operator pool index
          ($2n^2 - n + 1$ tokens for the full pool), plus coefficient
          grid tokens reused for $\gamma$ and $\beta$ parameters.
\end{itemize}

The problem tokens explicitly encode the interaction order and variable indices of the sub-problem. Each index token is followed by a coefficient token, so that the linear, quadratic, and cubic terms are represented directly in the input sequence. Thus, the token embeddings carry the detailed per-interaction information of the sub-problem, while the graph embedding provides only an additional global structural descriptor.

The coefficient tokens define a single shared numerical vocabulary that is used for two purposes. First, they represent the HUBO coefficients $c_i$, $c_{ij}$, and $c_{ijk}$ immediately following the corresponding problem index tokens. Second, the same vocabulary is reused in the circuit portion of the sequence to represent the ADAPT-QAOA parameters $\beta_k$ and $\gamma_k$ associated with each selected operator. Sharing one coefficient vocabulary for both problem coefficients and circuit parameters keeps the total vocabulary compact and allows the model to learn a common numerical embedding space, rather than maintaining separate vocabularies for problem values and circuit values.

A full token sequence takes the form:

\begin{equation}
\begin{aligned}
    &\texttt{<bos>},\;
      \underbrace{(i),\, c_i,\;\ldots}_{\text{linear}},\;
      \underbrace{(i,j),\, c_{ij},\;\ldots}_{\text{quadratic}},\\
    &\underbrace{(i,j,k),\, c_{ijk},\;\ldots}_{\text{cubic}},\;
      \texttt{<end\_of\_problem>},\\
    &\underbrace{\texttt{<new\_layer>},\; o_k,\; \beta_k,\; 
      \gamma_k,\;\ldots}_{L\text{ layers}},\;
      \texttt{<eos>},
\end{aligned}
\label{eq:token_seq}
\end{equation}

where $o_k$ is the pool index of the operator $O^{(k)} \in \mathcal{P}$ selected at layer $k$.

To provide additional structural information, a FEATHER embedding $\mathbf{e}_G \in \mathbb{R}^{500}$ is computed from the projected graph representation of each sub-problem and injected into the transformer input through a learned projection \cite{rozemberczki2020characteristic, tyagin2025qaoa}:
\begin{equation}
  X = E_{\text{tok}} + E_{\text{pos}} + W_{\text{emb}}\,\mathbf{e}_G
  \label{eq:gpt_input_hubo}
\end{equation}

where $E_{\mathrm{tok}} \in \mathbb{R}^{T \times d}$ and $E_{\mathrm{pos}} \in \mathbb{R}^{T \times d}$ denote the token and positional embedding matrices for an input sequence of length $T$, and $W_{\mathrm{emb}} \in \mathbb{R}^{d \times 500}$ is a learned projection matrix. The projected graph embedding $W_{\mathrm{emb}}\mathbf{e}_G$ is added to every token position as a fixed, sequence-independent conditioning signal. In this way, the token sequence provides the explicit interaction-level information, while the FEATHER embedding supplies a coarse summary of the overall structural shape of the sub-problem, analogous to class-conditioning embeddings used in conditional sequence models.

Training minimizes the standard cross-entropy next-token prediction loss. Separate GPT models should be trained for each sub-problem size.

\subsection{Inference: DQAOA-GPT Loop}

At inference, the trained model is used within a DQAOA loop to solve the large original optimization problem. Starting from an initial binary vector $\tilde{\mathbf{x}} \in {0,1}^N$, the algorithm iteratively samples $m$ sub-problems, generates optimized quantum circuits for each using the GPT model, executes the circuits on a quantum simulator or hardware to obtain candidate bitstrings, and aggregates the resulting local updates into the global solution.

For each sampled variable subset $V^{(k)}$, a sub-problem is constructed and encoded into the tokenized representation expected by the corresponding GPT model. The model then generates an adaptive quantum circuit for that sub-problem. After quantum circuit execution, the resulting bitstring $\mathbf{x}^{(k)}$ is used as a candidate local assignment. Each candidate update is accepted only if it decreases the energy of the original problem, ensuring monotonic improvement of the current solution.

Algorithm~\ref{alg:dqaoa_gpt} summarizes the full DQAOA-GPT procedure.

\begin{algorithm}
\caption{DQAOA-GPT for Combinatorial Optimization Problems}
\label{alg:dqaoa_gpt}
\begin{algorithmic}[1]
\REQUIRE $H$ (global problem), $N$ (global problem size), $m$ (number of sub-problems per iteration), $n$ (sub-problem size), $T$ (number of DQAOA iterations)
\ENSURE $\tilde{\mathbf{x}}$ (global solution)

\STATE Randomly generate $\tilde{\mathbf{x}} \in \{0,1\}^N$

\FOR{$t = 1$ to $T$}

    \FOR{$k = 1$ to $m$} 
        \STATE Randomly select $n$ variables $V^{(k)} \subseteq \{1,\ldots,N\}$
        \STATE Construct sub-problem $H^{(k)}$ using variables in $V^{(k)}$
    \ENDFOR

    \FOR{$k = 1$ to $m$}    
        \STATE $\triangleright$ \textit{Each sub-problem is independently solvable; parallelizable across GPUs}
        \STATE $\mathbf{x}^{(k)} \leftarrow \mathrm{QAOA\text{-}GPT}(H^{(k)})$
    \ENDFOR

    \FOR{$k = 1$ to $m$}
        \FOR{$j = 1$ to $n$}
            \STATE Replace $\tilde{x}_{V^{(k)}_j}$ with $x^{(k)}_j$ temporarily
            \IF{the updated solution reduces $H$}
                \STATE $\tilde{x}_{V^{(k)}_j} \leftarrow x^{(k)}_j$
            \ENDIF
        \ENDFOR
    \ENDFOR

\ENDFOR

\RETURN $\tilde{\mathbf{x}}$
\end{algorithmic}
\end{algorithm}

The aggregation step acts as a greedy coordinate-update procedure guided by GPT-generated local solutions. Because each accepted update strictly lowers the objective, the global energy is not increasing across iterations \cite{kim2024distributed, xu2025gpu}. At the same time, solving multiple sub-problems in parallel can enable efficient use of HPC resources and substantially accelerate the search over the high-dimensional solution space.

\begin{figure*}[!t]
    \centering
    \includegraphics[width=0.8\textwidth]{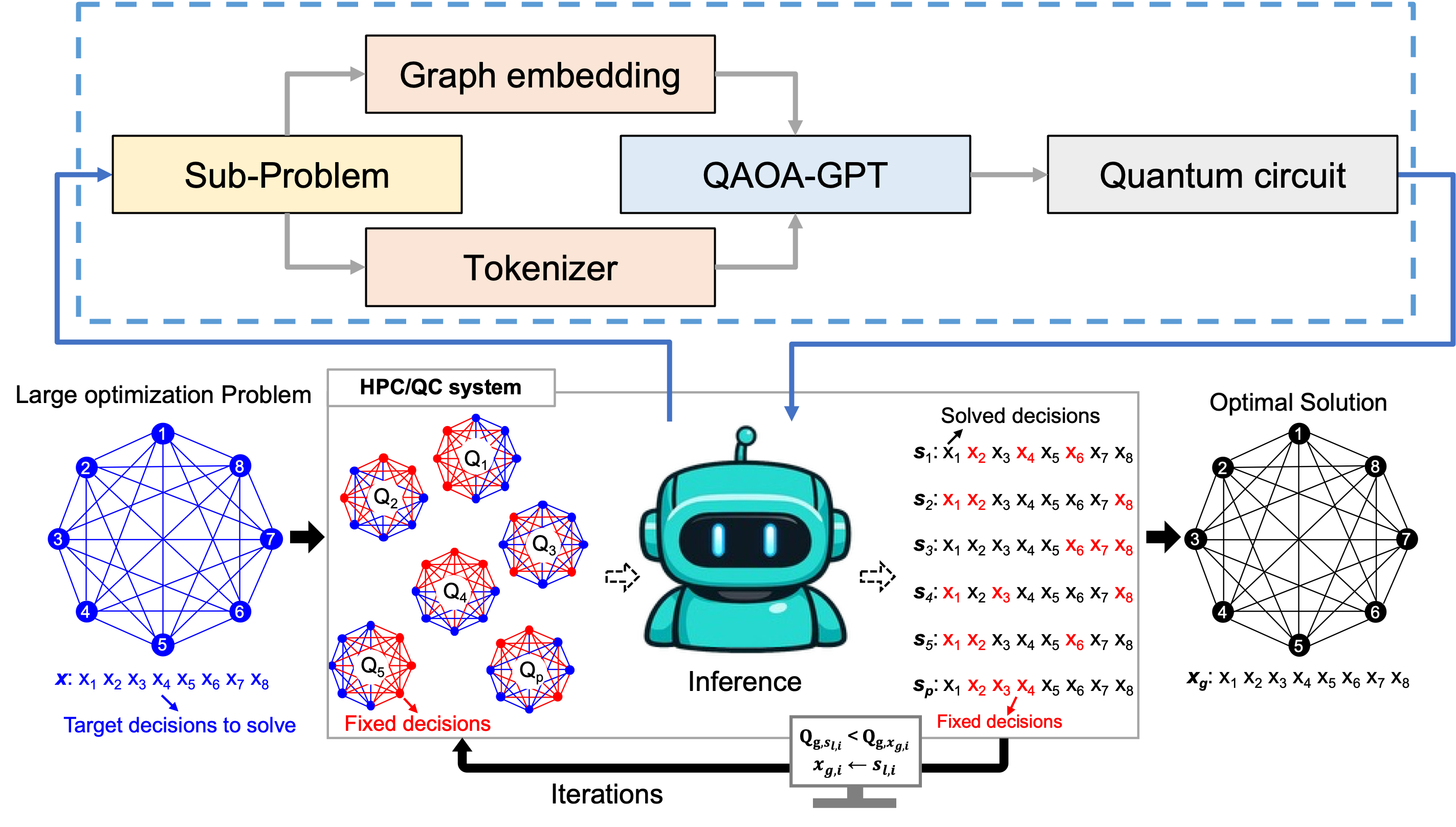}
    \caption{\label{fig:DQAOAGPT} Schematic illustration of the DQAOA-GPT workflow. Inspired by Refs.~\cite{kim2024distributed, tyagin2025qaoa}}
\end{figure*}

Figure~\ref{fig:DQAOAGPT} shows the overall DQAOA-GPT workflow.

\subsection{Experiments}

All computations are performed on the OLCF Defiant2 system, which consists of 20 HPE Cray XD220 CPU nodes and 2 HPE Cray XD670 GPU nodes. In this work, we primarily utilize the GPU nodes, each equipped with 8 NVIDIA H200 GPUs with 144 GB memory, connected via 400g InfiniBand interconnects. All quantum circuit simulations are carried out using the CUDA-Q SDK \cite{cuda_q} with the NVIDIA cuQuantum backend \cite{bayraktar2023cuQuantum}, providing a GPU-accelerated and consistent simulation environment for both DQAOA and DQAOA-GPT.

In this study, we use a single GPU to execute the DQAOA(-GPT) workflow, corresponding to $m=1$, where each sub-problem is processed sequentially. This setup is sufficient to demonstrate the overall framework and provides a fair comparison between DQAOA and DQAOA-GPT under identical computational conditions. For DQAOA-GPT, the reported runtime includes the full per-cycle inference pipeline, rather than only the execution time of the final accepted circuit. Specifically, the measured runtime includes graph projection, FEATHER embedding computation, GPT-based token generation, CUDA-Q simulation of all generated candidate circuits, and HUBO-energy evaluation of the resulting candidate bitstrings. For each sub-problem in each DQAOA-GPT cycle, we sample 10 candidate circuits using a sampling temperature of 0.8. The candidate producing the minimum-energy bitstring is then selected and accepted for the DQAOA update. Therefore, the reported DQAOA-GPT runtime reflects the full end-to-end cost of inference-based circuit generation and evaluation.

We note that this end-to-end runtime remains approximately constant as the sub-problem size increases because DQAOA-GPT replaces the iterative variational optimization loop with a finite number of generative inference and circuit-evaluation steps. In contrast, standard DQAOA requires repeated circuit evaluations and classical parameter updates, whose cost grows rapidly with the sub-problem size.

The proposed framework is also naturally parallelizable. In particular, multiple sub-problems within each DQAOA iteration can be solved independently, allowing straightforward distribution across multiple GPUs and nodes. Leveraging such parallel execution is expected to further reduce the overall runtime by enabling concurrent circuit generation and evaluation, providing a clear pathway for scalability in future studies \cite{kim2024distributed, xu2025gpu}.

\section{Results}

\begin{figure}[!ht]
    \centering
    \includegraphics[width=0.9\linewidth]{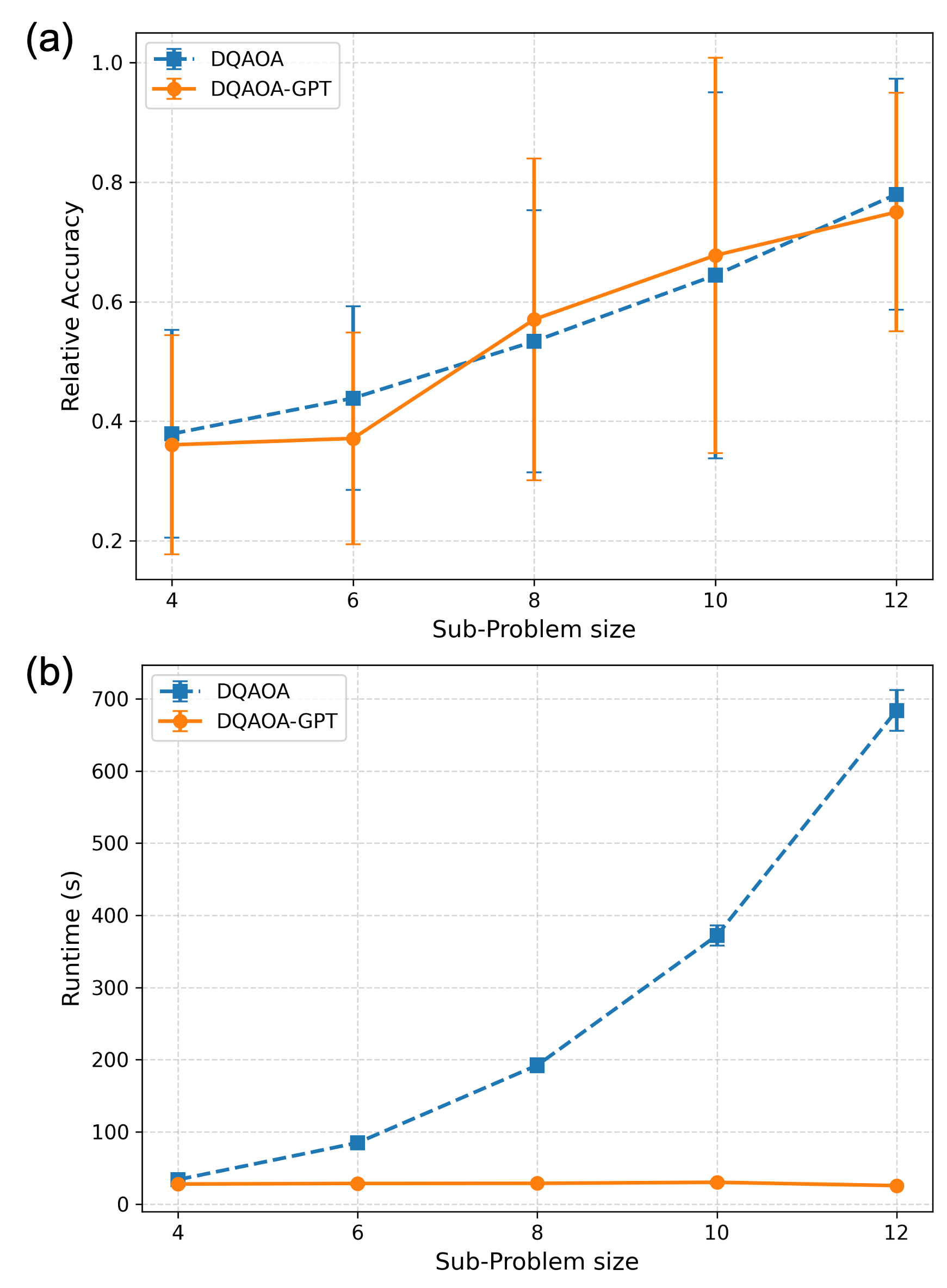}
    \caption{\label{fig:subPsize} DQAOA-GPT performance for different sub-problem sizes $n$. (a) Relative accuracy, and (b) runtime as a function of sub-problem size.}
\end{figure}

We investigate the effect of sub-problem size $n$ on DQAOA-GPT performance. Figure~\ref{fig:subPsize} shows relative accuracy and runtime as a function of $n$ for an original optimization problem of size $N=100$, with the number of DQAOA iterations fixed at $T=100$.
 
As sub-problem size increases, relative accuracy consistently improves. For $n=4$, relative accuracy remains low ($\sim$0.38 for DQAOA and $\sim$0.36 for DQAOA-GPT), whereas $n=12$ achieves up to $\sim$0.78, as shown in Fig.~\ref{fig:subPsize}(a). Larger sub-problems capture more of the higher-order interaction structure, leading to more effective local updates.

However, increasing $n$ substantially increases the computational cost of standard DQAOA. As shown in Fig.~\ref{fig:subPsize}(b), the runtime of DQAOA rises from $\sim$33.80 s at $n=4$ to $\sim$683.99 s at $n=12$, mainly due to the increasing cost of iterative variational optimization. In contrast, DQAOA-GPT maintains an approximately constant runtime of $\sim$28 s even as the sub-problem size increases, because the circuit generation step is replaced by a single inference process.
 
These results demonstrate that DQAOA-GPT greatly mitigates the accuracy--cost trade-off inherent in standard DQAOA, enabling the use of larger sub-problems to address larger optimization problems.
 
\section{Discussion}

We introduced DQAOA-GPT, a hybrid framework integrating distributed quantum optimization with generative circuit synthesis for large-scale combinatorial optimization problems. By replacing iterative variational optimization with GPT-based circuit generation, the framework eliminates the costly quantum-classical feedback loop, achieving substantial computational speedup. The advantage becomes more pronounced as sub-problem size increases, where variational optimization cost in DQAOA grows rapidly.

From a computational perspective, the proposed framework is inherently well-suited for HPC environments. Although the present study uses a single GPU to demonstrate the workflow and ensure a fair comparison, DQAOA-GPT naturally supports parallel execution across multiple GPUs and nodes, as sub-problem problems can be solved independently. This will provide a clear pathway for further runtime reduction and scalability through HPC-enabled parallelization.
 
Overall, this work highlights the potential of combining AI, HPC, and QC to address large-scale combinatorial optimization problems. By leveraging generative models for circuit synthesis, DQAOA-GPT offers a practical approach to overcoming key limitations of variational quantum algorithms. The proposed framework provides a scalable and efficient pathway toward hybrid quantum-classical optimization and represents a step toward demonstrating practical quantum utility in real-world applications.

\section*{Acknowledgment}
This research used resources of the Oak Ridge Leadership Computing Facility at the Oak Ridge National Laboratory, which is supported by the Office of Science of the U.S. Department of Energy under Contract No. DE-AC05-00OR22725. Portions of the text in this manuscript were refined with the assistance of the generative AI tool ChatGPT (OpenAI) to improve clarity and readability. In addition, a portion of Figure 2 was generated with the assistance of ChatGPT (OpenAI).

\printbibliography

\end{document}